\documentclass[conference,10pt]{IEEEtran}
\ifCLASSINFOpdf
\else
\fi
\usepackage{url}
\usepackage[pdftex]{graphicx}
\usepackage{algorithm}
\usepackage{listings}
\usepackage{courier}
\usepackage{comment}
\usepackage{booktabs}
\usepackage{multirow}

\usepackage{xcolor}

\definecolor{codegreen}{rgb}{0,0.6,0}
\definecolor{codegray}{rgb}{0.5,0.5,0.5}
\definecolor{codepurple}{rgb}{0.58,0,0.82}
\definecolor{backcolour}{rgb}{0.95,0.95,0.92}
\definecolor{codecyan}{rgb}{0.0,0.2,1.0}

\lstdefinestyle{mystyle}{
    commentstyle=\textcolor{codegreen},
    keywordstyle=\color{codecyan},
    numberstyle=\tiny\color{codegray},
    stringstyle=\color{codepurple},
    basicstyle=\ttfamily\footnotesize,
    breakatwhitespace=false,         
    breaklines=true,                 
    captionpos=b,                    
    keepspaces=true,                 
    numbers=left,                    
    numbersep=2pt,  
    firstnumber=auto,
    numberblanklines=false,
    showspaces=false,                
    showstringspaces=false,
    showtabs=false,                  
    tabsize=2
}

\newcommand{\vtkm}{VTK-m}
\newcommand{\vtk}{VTK}
\newcommand{\vtkmPN}{VTK-m-\textit{PN}}
\newcommand{\vtkmFM}{VTK-m-\textit{FM}}

\newcommand{\cppomp}{C++/OpenMP}
\newcommand{\knl}{KNL}

\newcommand{\ivybridge}{Ivy Bridge}

\begin{document}
\title{Performance Analysis of Traditional and Data-Parallel Primitive Implementations of Visualization and Analysis Kernels}

\author{\IEEEauthorblockN{E. Wes Bethel, David Camp, and Talita Perciano}
\IEEEauthorblockA{Lawrence Berkeley National Laboratory\\
Berkeley, CA, USA\\
Email: [ewbethel,dcamp,tperciano]@lbl.gov}
\and
\IEEEauthorblockN{Colleen Heinemann}
\IEEEauthorblockA{Lawrence Berkeley National Laboratory\\
Berkeley, CA, USA\\
University of Illinois \\
Urbana-Champaign, IL, USA \\
Email: cheinemann@lbl.gov}}


%Telephone: (800) 555--1212, Fax: (888) 555--1212}


\maketitle

% \fi
\IEEEpeerreviewmaketitle

\begin{abstract}

Measurements of absolute runtime are useful as a summary of performance when studying  
%parallel 
visualization and analysis methods on computational platforms of increasing concurrency and complexity.
We can obtain even more insights by measuring and examining more detailed measures from hardware performance counters, such as the number of instructions executed by an algorithm implemented in a particular way, the amount of data moved to/from memory, memory hierarchy utilization levels via cache hit/miss ratios, and so forth.
This work focuses on performance analysis on modern multi-core platforms of three different visualization and analysis kernels that are implemented in different ways: one is ``traditional'', using combinations of C++ and VTK, and the other uses a data-parallel approach using VTK-m. 
Our performance study consists of measurement and reporting of several different hardware performance counters on two different multi-core CPU platforms.
The results reveal interesting performance differences between these two different approaches for implementing these kernels, results that would not be apparent using runtime as the only metric. 
%\fix{(Any last minute changes needed here?)}
\end{abstract}


\section{Introduction}

%What is the problem we are trying to solve?
%comments about processor architecture complexity, programming models, and the desire for platform portability. 

Runtime, the measurement of elapsed time for a given method, is a commonly used metric in performance studies. 
While useful as a summary of performance, runtime by itself may not be enough information to reveal why a method performs the way it does.
At the other end of the spectrum, theoretical complexity analysis provides a useful understanding of the relationship between problem size $N$ and runtime, e.g., $O(N^2)$ vs. $O(N log N)$.
In between these two, theoretical analysis and measurement of absolute runtime, is an area where we can gain more insight about the actual performance of a method. 
Performance analysis in this regime often consists of collecting additional performance measures, such as hardware performance counters, which reveal much more information about algorithmic performance than either runtime or complexity analysis alone. 

Because computational platforms are rapidly evolving and increasing in complexity, a significant amount of research is underway to find pathways towards performance portability so that investments made in implementations perform well on a variety of different types of hardware now, and hopefully into the future.
%
%In the field of high performance visualization, the vast majority of such work uses runtime or runtime derivatives like frames/second as the basis for measuring performance improvements in algorithms and implementations.
%
%
This paper studies the measured performance differences that can result
%This paper explores the performance impacts that can result 
%on shared-memory platforms 
when using alternative implementations of key visualization and analysis kernels. 
%Why the problem is not already solved or other solutions are ineffective in one or more important ways
Most previous works in visualization performance analysis
%this space 
have focused on measuring and reporting only runtime. While runtime alone is a useful metric, deeper insights as to the nature of performance differences can result when considering additional metrics, such as hardware performance counters.

%Why our solution is worth considering and why is it effective in some way that others are not.
Our study focuses on three computational kernels common in visualization and analysis: isocontouring, particle advection, and stencil-based convolution. 
The stencil-based convolution is implemented using C++/OpenMP with \vtkm{} as the data-parallel alternative; the isocontouring and particle advection are written in C++ and invoke VTK methods as the traditional approach, and invoke \vtkm{} methods as the data-parallel alternative.
This study is motivated by the desire to compare off-the-shelf implementations of staple algorithms, with an eye towards deepening understanding of performance differences.

This study is timely because there is significant interest in having methods perform well on evolving platforms where there is increasing node-level concurrency, which is the consequence of the end of Dennard scaling~\cite{Dennard-Scaling:2011}. 
Some of the approaches for platform-portable parallelism include OpenMP-based loop parallelism~\cite{OpenMP:GPU:2018}, as well as \vtkm{}~\cite{Moreland:CGA2016}, where algorithms are constructed using sequences of data-parallel primitives.
The visualization community has a keen interest in \vtkm{}, and most of the performance studies this far have used only runtime (and derivatives) as the performance metric.
What our study adds is deeper insight and understanding of why these algorithms perform the way they do.
%
%We explore alternative shared-memory implementations, which include OpenMP-based loop parallelism~\cite{OpenMP:GPU:2018}, as well as VTK-m~\cite{Moreland:CGA2016}, where algorithms are constructed using sequences of data-parallel primitives.
%
%While distributed-memory parallelism is a topic of interest and active research, we are focusing on shared-memory parallelism in this work due to the technology evolution occurring at the node level, including the need for platform-portable approaches for algorithmic development.~\fix{need citations here to back this up.}

%What is contribution of the paper.
The contributions of this work include: %\fix{(revise this contributions list, make it emphatic!)}
\begin{itemize}
    \item A first-of-its kind  analysis using hardware performance counters comparing traditional (\vtk{} and C++) and emerging (\vtkm{}, data parallel primitives) implementations of staple methods: nearly all previous works focus exclusively on using runtime as the performance measure. 
    \item Deeper insights into the types of performance impacts that can result from reformulating algorithm design using data-parallel primitives.
    \item The performance study methodology is straightforward and readily applicable to a much broader set of applications, and is particularly relevant given research interest in evolving methods to increasingly complex and heterogeneous hardware. 
%        \item An in-depth performance analysis of three staple visualization and analysis kernels comparing traditional and more novel (VTK-m) implementations.
    \item Two new and different \vtkm{} implementations of a stencil-based computation, with a comparative performance analysis of both approaches that provides some insight into performance characteristics of different types of \vtkm{} worklets. 
\end{itemize}
\section{Background and Previous Work}

%\subsection{Parallel Visualization and Analysis}
\subsection{The Evolving Computational Landscape of Visualization and Analysis}
%\fix{fixme}

%Over the years, processor and platform architectures have evolved significantly, as have the software environments and programming languages used by developers.

%Quick whirlwind tour of how people have done parallelization over the years. 

%Talk about how the parallel framework reflects what is needed to run on a given architecture. MPPs in the 1990s gave rise to MPI. Threads, and later, OpenMP for running on SMPs.

%\fix{Need citations in this paragraph.}
In response to the end of Dennard scaling, system and processor architectures have evolved to use deepening memory hierarchies combined with increasing node-level parallelism~\cite{Dennard-Scaling:2011}.
Over the past decade or so, numerous programming models and environments --- such as OpenMP~\cite{chandra2000parallel}, OpenCL~\cite{OpenCL:2011}, and Kokkos~\cite{Kokkos:2014}, to name a few --- have emerged where the objective is to provide for platform portability as well as efficient shared-memory parallelism.

For distributed-memory parallelism, codes like VisIt~\cite{Childs:VisIt-HPV-Chapter:2012} and ParaView~\cite{Paraview} use the Message Passing Interface (MPI)~\cite{Walker96mpi} as the programming model to achieve platform portability. These tools each leverage VTK's core algorithms~\cite{vtk-book:1998}, and manage their execution in slightly different ways that are implementation specific. 

In the visualization community, the VTK-m library~\cite{Moreland:CGA2016} follows a similar technology trajectory, where user code can be executed with one of several different device- or platform-level backends, such as TBB~\cite{tbb-book}, OpenMP~\cite{chandra2000parallel}, or CUDA~\cite{Nvidia-ProgramGuide}.
\vtkm{} is positioned to be a follow-on to the VTK library~\cite{vtk-book:1998}, which has served as a stalwart in the visualization community for more than two decades but is limited to predominantly serial use due to a combination of factors, including close entanglement of data structures and execution models and use of static variables that maintain state and are hence not thread-safe. 

Given the community interest in \vtkm{}, our work here seeks to provide more insight into the performance differences that result when implementing algorithms using traditional, either straight C++ or C++/VTK, with emerging approaches like \vtkm{} that offer platform portability using data parallel primitives (DPPs) as the algorithmic building block.

\subsection{Comparing Traditional and VTK-m Implementations}

%\fix{fixme}

%Cite our own work where we've done runtime comparison of different graph algorithms done in OpenMP, threads, VTKm.

Several recent works have focused on developing \vtkm{} implementations of key algorithms and comparing them to custom, traditional implementations.
Larsen, et al. 2015~\cite{RayTracing:2015} study the performance of a raytracing renderer implemented using data-parallel primitives with two reference implementations. Their reported metric is frame/second, which is a derivative of runtime. 
Pugmire, et al. 2018~\cite{PugmirePGV18} present a 
%data-parallel design and 
VTK-m implementation of particle advection. 
%In addition to measuring strong scaling characteristics, 
They compare the performance of their implementation, which uses a \emph{parallelize over seeds} approach, against two others: a hand-coded implementation, and the VisIt~\cite{Childs:VisIt-HPV-Chapter:2012} application. All their studies rely on measurements of runtime. 

In the area of graph algorithms, Lessley, et al., 2017~\cite{Lessley:MCE-DPP:LDAV:2017} describe a VTK-m implementation of a method for maximal clique enumeration and compare its performance with reference implementations on CPU and GPU platforms. 
Lessley, et al., 2018~\cite{Lessley:DPP-PMRF:LDAV:2018} present a VTK-m implementation of a method for probabilistic graphical modeling optimization, a form of unsupervised learning, and compare its performance to a reference implementation on a CPU. In both these works, the basis for comparison is elapsed runtime. 

Recent work by Perciano, et al., 2020~\cite{Perciano:2020:ISC} delves more deeply into use of LIKWID and performance counters to gain deeper insight into performance differences between three different shared-memory parallel implementations of an unsupervised learning method using graphical model optimization: C++/pthreads, C++/OpenMP, and \vtkm{}. 
That study's finding is that the \vtkm{} algorithmic reformulation  executes far fewer instructions than the traditional counterpart, owing to the significant algorithmic refactorization that occurs going from traditional coding style to one that uses data parallel primitives.
Our work is similar in that we are diving deep into the collection and analysis of performance counters, and using them to shed insight into the performance differences between traditional and \vtkm{} implementations.
However, we are focusing on comparing staple visualization kernels rather than graphical model optimization. 

%Yenpure, et al. 2019~\cite{Yenpure:EGPGV19} describe a data-parallel implementation of a point-merging operation in VTK-m. They compare four variants of a data-parallel VTK-m implementation with two variants of a VTK implementation, and report a metric \emph{processing rate} that measures number of points processed per second, a metric that is a derivative of runtime. 

One primary difference between these previous works, except for Perciano, et al., 2020~\cite{Perciano:2020:ISC}, and our work here is the deeper introspection provided by using detailed hardware performance counters.
These additional metrics offer the ability to better understand why a given code performs better or worse in a particular set of circumstances, and also helps to provide a more sound basis for performance analysis. 

%These don't take into account hardware performance counters.

%\subsection{Performance Analysis Methodologies}

%Cite some recent examples showing the types of metrics that people have used. Including VTK-m based methods. If I'm not mistaken, the bulk of these studies have not really focused on comparing the VTK-m version with a non-vtkm version.

%These don't take into account hardware performance counters.

\section{Design and Implementation}

The focus of this section is on presenting concepts related to design and implementation using ``traditional'' and data-parallel approaches. 
In the case of isocontouring (\S\ref{sec:design-isocontour}) and particle advection (\S\ref{sec:design-padvect}), we briefly describe the implementations used in our study. The core VTK and \vtkm{} implementations existed already as part of previous work done by others.
In the case of stencil-based convolution (\S\ref{sec:design-stencil}), we describe our traditional and \vtkm{} implementation in some detail, as this is new work that did not exist prior to this study.

A comment about parallelism: while \vtkm{} is parallel capable by its very design, VTK is not.
Two of the three methods we present (isocontouring, particle advection) use VTK implementations for the ``traditional'' approach: we created applications that load data and invoke the appropriate VTK (or \vtkm{}) method. 
The VTK implementation of these methods themselves are not parallel, nor are they readily parallel capable: due to the existence of thread-unsafe constructs, such as static variables inside VTK that hold state, they cannot be invoked in parallel by multiple concurrent execution threads.

%The general recipe in this section is to first describe the generic algorithm, then describe the parallel implementations.

\subsection{Isocontouring}
\label{sec:design-isocontour}

% Describe isocontouring.
Isocontouring in 2D and 3D is a staple visualization method because it is useful for showing geometric structure associated with specific data values.
The seminal method for computing isosurfaces is from Lorenson and Cline, 1987~\cite{Lorensen87marchingcubes:}. 

This algorithm operates on scalar fields by considering the cell formed by eight neighboring points that form a hexahedron. For each such cell, the algorithm performs a classification step on each vertex of the cell, where the classification is the union of booleans indicating if the scalar value at each node is above or below the contouring value.
Then, the algorithm uses this classification as an index into a lookup table containing a specification of a triangulation of the surface passing through a cell with that given configuration: some configurations have one triangle, others may have two or more triangles.
Then, the algorithm computes isosurface triangle vertex values using an interpolant, typically trilinear interpolation. 
The triangles from all such computations at each cell comprise the resulting isosurface.

It is well beyond the scope of this paper to enumerate the improvements and enhancements over more than 30 years since this algorithm first appeared in 1987.
In brief, some of the key improvements consist of resolving some of the topological ambiguities that arose in the original formulation~\cite{nielson:1991}, and
use of spatial data structures to accelerate finding cells that contain the isosurface (c.f., Newman and Yi, 2006~\cite{Newman:2006}). 

For the purpose of our study, we are interested in using and benchmarking two reference implementations: the \texttt{vtkContourFilter} method from VTK, and the \texttt{vtkm::filter::Contour} method in VTK-m, which is a reimplementation of the \texttt{vtkContourFilter}~\cite{vtkmContourFilterWeb:2020}.
Both the VTK and VTK-m methods perform the same fundamental algorithmic processing steps --- cell classification and triangle generation --- but use different mechanisms for doing so. 

The VTK-m implementation uses a combinations of \texttt{worklets}, which are instantiations of data parallel primitives. 
For example, it uses a \texttt{WorkletVisitPointsWithCells} for the \texttt{ClassifyCell} operation. This type of worklet will ``apply a function (the operator in the worklet) on all elements of a particular
type (such as points or cells) and creates a new field for those elements''~\cite{vtkm-userguide-1.5}, with the added feature that nearby, incident cells are also accessible. 
It also uses other types of worklets, such as \texttt{ScatterPermutation} and \texttt{ScatterCounting}, which are used to create multiple outputs (each of which satisfies some condition) from an input, as well as others.
Both VTK and \vtkm{} implementations are sophisticated works of complex design and engineering.

For this study, we created two applications, each of which loads data and writes output.  One version invokes VTK's \texttt{vtkContourFilter}, and the other invokes VTK-m's \texttt{vtkm::filter::Contour} method.
One primary difference between the VTK and VTK-m implementations concerns parallelism. The VTK-m implementation, through its execution environment, is capable of running worklets in shared-memory parallel fashion using a number of different device backends (e.g., OpenMP, TBB, CUDA, etc.). The VTK-m device execution environment will decompose a worklet's workload into chunks that execute in parallel. 
This type of decomposition and parallel operation does not exist in much of VTK, including the \texttt{vtkContourFilter}, which is limited to serial use only.

\subsection{Particle Advection}
\label{sec:design-padvect}

Particle advection is a technique for calculating the trajectory a particle follows through a flow field.
An integral curve –- commonly referred to as a pathline –- encodes the trajectory
of a single massless particle, which in turn gives insight into
the flow behavior in the area surrounding the particle’s path.

%If $x$ is a spatial location of a point and $t$ is a time,
%then the vector field $V$ maps the tuple $(x,t)$ to its velocity, as $V(x,t)$.

Advection constructs integral curves, which are continuous functions
tangential to the vector field. The curves are solutions to an ordinary
differential equation, and, for an integral curve $I$ and vector field $V$, can be represented
as:
\begin{equation}                                                                
{d \over dt} I(t) =  v(I(t),t)
\label{eqn:padvect}
\end{equation}    
 where $I(t_0) = x_0$, and for a seed point at location $x_0$ at time $t_0$ ~\cite{Agranovsky:LDAV2014}.
%

%Describe particle advection in general. Just enough to say what it is, and mention a few previous works.

As with the isocontouring study, our focus is to compare performance of two existing reference implementations, one from VTK and the other from VTK-m.
From VTK, we are using the \texttt{vtkStreamTracer} method, and from VTK-m, we are using the \texttt{vtkm::filter::Streamline} method. Both methods take as input a vector field and then perform integral curve calculation to compute streamlines from a set of input seed points with a user-specified integrator (e.g., Runge-Kutta4). 
Our application that uses the VTK implementation of the \texttt{vtkStreamTracer} method is serial in nature and cannot be trivially parallelized due to explicit limitations in the VTK implementation.

From \vtkm{}, the \texttt{vtkm::filter::Streamline}
 implementation uses a parallelize-over-seeds approach, which is described in Pugmire, et al., 2018~\cite{PugmirePGV18}.
Their implementation uses a \texttt{Field Map} worklet, where the worklet is invoked at each input datum, which in this case, consists of a list of seed points. VTK-m's internal execution model will parallelize this operation by invoking the worklet in parallel, where each invocation will process some number of input seed locations.

%Our application uses an OpenMP implementation that will invoke the  \texttt{vtkStreamTracer} method in parallel: if there are $N$ seed locations and $P$ OpenMP threads, then each thread will be asked to process $N/P$ seeds in parallel. 

%which is a filter that integrates a vector field to generate streamlines~\cite{vtkStreamTracerWeb:2020}. 
%, which performs the same

%Over the years, there has been work to explore different parallelization strategies.

%Non-VTK-m: is there a parallelize over seeds implementation in VTK? Or can you grab/use the "specialized implementation" one described in Pugmire's 2018 EGPGV paper?~\cite{PugmirePGV18}.

%VTK-m implementation: Use Pugmire's parallelize-over-seeds implementation, which is described in their EGPGV 2018 paper~\cite{PugmirePGV18}, and which is hopefully part of the VTK-m distro.

%These codes have unstructured memory access patterns.

\subsection{Stencil-based Convolution}
\label{sec:design-stencil}

In a stencil based computation, each point of a multidimensional grid is updated with contributions from its neighbors. This form of computation lies at the heart of many different types of scientific computations, such as solving partial differential equations on a regular, structured grid (c.f.,~\cite{Roth97compilingstencils}). 
We include this computationl pattern because it is common in many types of computing applications, such as numerical simulation, image analysis/computer vision, convolutional neural networks, and more.
%\fix{Say something about why we are focusing on this problem.}

In our study, we focus on a particular type of stencil-based computation, namely image convolution.
This form of computation is a structured memory access code, where memory is accessed in a regular and predictable fashion. 
In particular, our computation performs Guassian smoothing, which is a spatial image processing filter.
%
%
%specific stencil-based operation: Gaussian smoothing. While it is a relatively simple stencil operation, stencil-based operations play a significant role in many types of operations in scientific computing and image processing.
%
%
In this computation, each destination pixel $d(i)$ is a sum of nearby pixels averaged using a weighting scheme that gives more weight to pixels closer to $i$, and less pixels further away (Eq.~\ref{eqn:gaussianSum}).
\begin{equation}                                                                    
d(i) =  \sum{g(i, \bar{i})}
\label{eqn:gaussianSum}
\end{equation}         
 where the Gaussian weights are given by
\begin{equation}
g(i,\bar{i}) = e ^ { -{\frac{1}{2}} \left( { \frac{{\delta(i,\bar{i})}}{ \sigma_d } } \right) ^2 }
\label{eqn:distanceFunc}
\end{equation}  
In Eq.~\ref{eqn:distanceFunc},
$\delta(i,\bar{i})$ is the distance between pixels $i$ and $\bar{i}$. 
$\sigma$ is a parameter that defines whether the weights are more tightly focused around the source pixel (somewhat less smoothing), or if the weights are more diffuse, and give more weight to pixels further away (somewhat more smoothing).

In practice, this type of computation is bounded such that a stencil of size $R^N$, in $N$ dimensions, contains filter weights computed such that the integral of Eq.~\ref{eqn:distanceFunc} over the set of $R^N$ weights sums to 1.
Then, computing the smoothed pixel value consists of performing a sum of products of the source pixel weights with the 
%Gaussian 
filter weights. An example for 2D image convolution appears in Listing~\ref{listing:stencil-core}, and this computation trivially generalizes to $N$ dimensions. 

\begin{lstlisting}[caption={Stencil computation in 2D: performs sum of product of nearby pixels with weights.},label={listing:stencil-core}, name=stencil-core, float=t, style=mystyle,language=C++]
float smoothPixel(Si, Sj, S, R, weights)
{
    // compute the weight sum of pixels nearby
    // this code doesn't handle edge conditions
    // and assumes sum of weights[i,j] = 1.0
    float sum = 0.0;
    for (int j=0; j<R; j++)
        for (int i=0; i<R; i++)
           sum += weights[i,j]*S[Si+i,Sj+j]
    return sum;
}
\end{lstlisting}

An application that uses the \texttt{smoothPixel} method will iterate over the pixels/voxels in a source image/volume, and invoke this method at each pixel/voxel.
This type of application is straightforward to parallelize in that the sense that the computation at each $d(i)$ is independent of the computation at all other locations; these computations may be performed independently and in parallel. 

Listing~\ref{listing:stencil-openmp} shows one such parallelization using OpenMP. In that listing, each of the OpenMP threads is assigned a row (scanline) of pixels to work on.
OpenMP does assignments using one of several different strategies, such as round-robin, etc. depending on the setting of a runtime environment variable (c.f.~\cite{chandra2000parallel}).
This type of parallelization is relatively coarse-grained in the sense that each thread is assigned a significant amount of work.
 
%\begin{lstlisting}[caption={foo},label={listing:stencil-openmp}]
\begin{lstlisting}[caption={OpenMP parallel computation: each thread gets a scanline to process.},label={listing:stencil-openmp},name=stencil-openmp, float=h!,style=mystyle,language=C++]]
#pragma omp parallel for
for (j=0;j<imageHeight;j++)
   for (i=0;i<imageWidth;i++)
       destImage[i,j] = smoothPixel(i, j, S, R, weights)
\end{lstlisting}
 
One potential VTK-m implementation of this stencil operation is to use the same computational kernel shown in Listing~\ref{listing:stencil-core}, but then let VTK-m manage how this computation is invoked.
The basic concepts are as follows. First, we define a Field Map worklet that has input and output parameters that are ``arrays''; these are essentially \texttt{std::vector} objects. Then, after populating the input array with the source image, we invoke the dispatcher that invokes the \texttt{ImageConvolutionWorklet}, as shown in Listing~\ref{listing:stencil-vtkm-fm}. 
Then, VTK-m will invoke that worklet once per array item using one of several different potential device backends, depending upon user build configuration options and runtime choices (see the VTK-m User's Manual for more details~\cite{vtkm-userguide-1.5}). 
In the case of our application, this worklet is invoked once per input pixel. The result is a much finer-grained approach to parallelism than the scanline-based parallelism shown in Listing~\ref{listing:stencil-openmp}.
We refer to this design as the \vtkmFM{} method, where \textit{FM} refers to the use of the \textit{Field Map Worklet}, and which here uses an explicit indexing computation to access specific locations in the input and output image data arrays.
%refers to the explicit indexing computation we must perform to access specific locations in the input and output image data arrays. 

% \fix{this explanation of VTK-m code doesn't suffice, it needs to be improved.}
 
\begin{lstlisting}[caption={VTKm-\textit{FM} algorithm: In VTK-m, execution environment iterates over the field and invokes the smoothPixel worklet in parallel.},label={listing:stencil-vtkm-fm}, float=h!, name=stencil-vtkm-fm, style=mystyle,language=C++]]
template <typename InputArrayType, typename OutputArrayType>
	VTKM_EXEC void operator()(const InputArrayType & inputImg, OutputArrayType & outputImg, 
	vtkm::Id indx) const
	{
	    /// assume private R: stencil size
	    /// explicit indexing: compute (i,j) indices from 1D indx
	    int jVal = indx / nCols - R/2; // which row
	    int iVal = indx % nCols - R/2; // which column
	    
	    float sum = 0.0;
	    for(int j=0; j<R; j++)
	       for(int 0; i<R; i++)
	          sum += weights[i,j]*inputImg[iVal+i,jVal+j];
	    outputImg.Set(indx, sum);
	}
...
/// then invoke it in main():
vtkm::worklet::DispatcherMapField<vtkm::worklet::ImageConvolutionWorklet> dispatcher(myWorklet);
dispatcher.Invoke(inputImageArray, outputImageArray);
\end{lstlisting}
 
There are other potential implementations of this stencil that we could pursue in \vtkm{}, implementations that could potentially take greater advantage of VTK-m's data parallel primitives (DPPs). 
%Those DPPs are implemented as ``Device Algorithms'' 
These DPPs include operations like Reduce, Sort, CopyIf, ScanExclusiveByKey, and so forth.
Some interesting future work would entail exploring recasting the stencil computation in terms of using VTK-m's DPP Device Algorithms, as has been done with other types of computations, such as probabilistic graphical modeling optimization ~\cite{Lessley:DPP-PMRF:LDAV:2018}. 

We did explore a second VTK-m implementation, namely one that uses a \texttt{Point Neighborhood Worklet}, which we refer to as the \vtkmPN{} algorithm.
Like the \vtkmFM{} algorithm, \vtkm{} invokes the worklet once per input grid location.
One significant difference is that the \vtkmPN{} algorithm may access
field values of nearby points within a neighborhood of a given size,  as opposed to having access to the entire mesh. 
In other words, the \vtkmFM{} method needs to do its own indexing: it is handed a 1D input index, and then needs to use this input to produce an index into a multidimensional array.
In contrast, methods that use the \texttt{Point Neighborhood Worklet} will use an \texttt{inputData.Get()} method to access data, rather than using an index computation.
Our \vtkmPN{} algorithm is shown in Listing~\ref{listing:stencil-vtkm-pn}.
 
 \begin{lstlisting}[caption={\vtkmPN{} algorithm: similar to the \vtkmFM{} algorithm, but without the global indexing computation as VTKm provides a view only to the local mesh/image neighborhood},label={listing:stencil-vtkm-pn}, float=h!, name=stencil-vtkm-pn, style=mystyle,language=C++]]
  template < typename InputFieldPortalType >                                                             
    VTKM_EXEC typename InputFieldPortalType :: ValueType operator ()(
    const vtkm :: exec :: FieldNeighborhood < InputFieldPortalType >& inputField ,                       
    const vtkm :: exec :: BoundaryState & boundary ) const                                               
    {  
        auto minIndices = boundary.MinNeighborIndices (this -> stencilRadius );
        auto maxIndices = boundary.MaxNeighborIndices (this -> stencilRadius );
        
        float sum=0.0;
	    for(vtkm::IdComponent j=minIndices [1]; j<=maxIndices[1]; ++j)
           for (vtkm::IdComponent i=minIndices[0]; i<=maxIndices[]0]; ++i)
              sum += inputField.Get(i, j) * this->filterWeights[f(i,j)];                    
        
        return static_cast <T> (sum);     
	}
...
/// then in main():
// set the image dimension sizes
vtkm::cont::CellSetStructured<2> myGrid2D
myGrid2D.SetPointDimensions(vtkm::Id2(nCols, nRows);

// create the dispatcher and invoke it
vtkm::worklet::DispatcherPointNeightborhood<vtkm::worklet::ImageConvolutionWorklet> dispatcher(myWorklet);
dispatcher.Invoke(myGrid2, inputImageArray, outputImageArray);
\end{lstlisting}

%\end{lstlisting}

\section{Results}
%This is a section.

\subsection{Research Questions}

%Enumerate the research questions we are trying to answer.
One of the primary areas of study is to better understand the key performance characteristics of a ``traditional'' implementation with an implementation that uses \vtkm{}s data-parallel primitives (DPPs).
While we do report runtime, we collect and study hardware performance counter data to gain deeper insight into the potential factors contributing to runtime differences. 

Given that the \vtk{} implementations of isocontouring and particle advection are serial, we focus our tests on comparing serial \vtk{} and \vtkm{} implementations. 
While these \vtkm{} implementations do execute in parallel, we are not performing any scaling comparisons here between the \vtk{} and \vtkm{} implementations of isocontouring and particle advection.
In contrast, the stencil computation runs in parallel in all configurations, and so we include a scaling study for that method.

Given these limitations, our primary objective is to gather hardware performance counters and analyze the results to better understand the reasons for runtime performance differences between \vtk{} and \vtkm{} implementations.

\begin{comment}

One of the primary research questions is to better understand the  the key performance characteristics of C++/OpenMP and VTK-m/OpenMP algorithms? For a given problem, what differences do they show in terms of scalability, in terms of memory system utilization, in terms of vectorization, and so forth.

For each of the three different test configurations -- isocontouring, particle advection, stencil-based convolution -- 
 we first look at runtime of the serial version of each of the methods. Does one implementation execute more quickly than another? And then dig more deeply into performance counters to look for reasons why.

Second, we look at scalability of methods with a relatively narrow set of constraints: strong scaling on a single socket of two different multi-core platforms. Using the insights from the study of serial performance analysis, what additional insights from performance counters shed light on scaling characteristics?

\end{comment}

\subsection{Methodology}

%Then, the methodology explains how we will set up and run the experiment to get answers to the question.

\subsubsection{Computational Platforms}

% from the PMRF IPDPS 2020 submission:
\emph{Intel Xeon Phi and Intel Haswell.} \texttt{Cori.nersc.gov} is a Cray XC40 system comprised of 2,388 nodes containing two 2.3 Ghz 16-core Intel Haswell processors and 128 GB DDR4 2133 MHz memory, and 9,688 nodes containing a single 68-core 1.4 GHz Intel Xeon Phi 7250 (Knights Landing) processor and 96 GB DDR4 2400 GHz memory. For our experiments, we use the KNL processor, where each core  has a 32 KiB L1 cache, and each pair of cores share a 1 MiB L2 cache (L2 is the last level cache on this platform).
%\footnote{Cori configuration page: \url{http://www.nersc.gov/users/computational-systems/cori/configuration/}}.
On this platform we are using Intel's C/C++ compiler, icc (ICC) 19.0.3.199 20190206. Compiler flags for both \vtk{} and \vtkm{} include the following optimization options to enable vectorization:
\texttt{-O3 -march=knl -mtune=knl -DNDEBUG -funroll-loops}.
%\fix{(Need to say if we are using Haswell nodes, KNL nodes, or both.)}

\emph{IvyBridge.} \texttt{Allen.lbl.gov} is an Intel(R) Xeon(R) CPU E5-2609 v2 containing 2 2.5 GHz 4-core Intel Xeon IvyBridge EN/EP/EX processors and 32 GB of memory. 
Each core has a 32 KiB L1 cache, a 256 KiB L2 cache, and all cores share a 10 MiB L3 cache (L3 is the last level cache on this platform).
On this platform, we are using Intel's C/C++ compiler, icc (ICC) 19.1.0.166 20191121.
Compiler flags for both \vtk{} and \vtkm{} include the following optimization options to enable vectorization:
\texttt{-O3 -march=ivybridge -mtune=ivybridge -funroll-loops}.

\subsubsection{Software Environment}

\textit{VTK and VTK-m.}
We are using VTK version 8.2.0, and VTK-m version 1.5.0. 
For all tests, we use VTK-m's OpenMP backend, which generates OpenMP shared-memory parallel code that runs on multi-core CPU platforms.
While VTK-m is capable of emitting CUDA-based code, and while modern versions of OpenMP are also capable of emitting CUDA-based device code, we did not include GPU as one of the platforms in this study; this will make for interesting future work.
%\fix{Need appropriate citations.}

% compiler flags for VTK on cori: CXX_FLAGS = -O3 -march=knl -mtune=knl -DNDEBUG -funroll-loops  -O3 -DNDEBUG -fPIC -fvisibility=hidden   -std=c++11
% compiler flags for VTK-m on cori: CXX_FLAGS =  -Wall -pipe -O3  -march=knl -mtune=knl -funroll-loops  -DNDEBUG -fPIE   -ffunction-sections -qopenmp -std=c++11

%What datasets, implementations/versions (VTK, VTK-m, compilers), ...

\subsubsection{Data Sets and Algorithmic Parameters}
\label{sec:study-data}
%For the isocontouring, what dataset are we using?

%For the particle advection, what dataset are we using?

For the stencil-based smoothing and isocontouring studies, we are using a scientific dataset that was obtained by the Lawrence Berkeley National Laboratory Advanced Light Source X-ray beamline 8.3.2
%~\footnote{\url{microct.lbl.gov}}
~\cite{Donatelli:2015}.
This dataset contains cross-sections of a geological sample and conveys information regarding the x-ray attenuation and density of the scanned material as a gray scale value. 
The original data consists of a stack of 500 images at resolution $1290\times 1305$.

For the stencil-based smoothing study, we are using an augmented version of the dataset where we use one 2D slice, and then replicate it twice in X and Y to produce an image of resolution $5160 \times 5220$. 
The size of the convolution kernel is $19 \times 19$ pixels, and $\sigma = 0.33$.

For the isocontouring study, we use a $400^3$ subset of a processed version of the original $1290\times1305\times500$ volume. The processing consists of multiple stages in an image analysis pipeline, where we use a custom, high-quality image segmentation algorithm based on an unsupervised learning method~\cite{Lessley:DPP-PMRF:LDAV:2018}, and then run the segmented image through a 3D version of the stencil-based smoothing code presented in this study. We performed contouring with an isocontour level of 15, which results in the surface that appears in Fig.~\ref{fig:sand400-iso15}.
Note that since we are focusing on results from serial runs only, parallel performance variations that might result from various conditions such as data-dependent factors or domain decomposition effects are not of concern in these studies.
%This dataset is a good candidate for our isocontouring study because the spatial distribution of cells containing the isocontour is relatively even throughout the volume, which will help to minimize the potential for load imabalance.

%For the particle advection study, we are using three different datasets, each with different characteristics that will exercise different aspects of the particle advection algorithm's data-dependent behavior.

For the particle advection study, we are using a vector field data set produced by 
%A fusion dataset comes from 
the NIMROD code, which is performing numerical modeling of a  magnetically confined plasma in a tokamak device~\cite{SOVINEC2004355}. 
The magnetic field lines, through which we are doing particle advection, travel around and around the toroidal-shaped domain in a helical fashion. As a result, most particles will not exit the domain, and the amount of work for each seed point will be about the same over the lifetime of the integration.
For both \vtk{} and \vtkm{} runs, we use 500 seed points placed equidistantly along a diagonal running through the domain, and compute particle positions for 1000 steps using the Runge-Kutte4 integrator. 
Note that since we are showing only serial \vtkm{} results, parallel performance variations as a function of seed point placement are not of concern in these studies.

\textit{Hardware Performance Counters.} For this study, we are leveraging the LIKWID software infrastructure v4.3.4~\cite{psti:2010, LIKWID-website:2020}.
LIKWID is a set of lightweight, command-line tools that are useful for obtaining measurements of hardware performance counters on Linux platforms in user space.
While LIKWID is capable of collecting performance counters for an unmodified application, we make use of LIKWID's Marker API to collect performance data only in the section of code containing the computational algorithm; we do not include other operations, like data I/O, in the performance counter data.
Table~\ref{table:perfCounters} describes the performance counters we gather for analysis in all the experiments.

\begin{table}[t!]
\footnotesize
    \centering
       \caption{Hardware performance counters and other measures we use in our experiments. }
    %\fix{this table needs to be updated, and possibly condensed. }}
    \label{table:perfCounters}
    \begin{tabular}{p{1.0in}p{2.0in}}
    Performance counter, measure & (Source) Description \\
    \hline
 \texttt{INSTR\_RETIRED\_ANY} & (LIKWID) Shows how many instructions were completely executed, and does not include speculative instruction loads~\cite{instRetiredEventWeb:2020}. \\
    \\
\texttt{FLOPS, FLOPS\_DP, FLOPS\_SP} & (LIKWID) Count of the number of single- and double-precision floating point operations that were executed. On the KNL platform, there is no way to differentiate between single- and double-precision operations. \\
%    On the Ivy Bridge, there is the known potential for inaccuracies with these measures, whereby the counters may include both instructions issued and retired. For the purposes of the test results we show here, we assume some ``error bars`` on the results, but consider the results informative within the broader context that considers counts of both FLOPs and other instructions executed. \\
    \\
Cycles Per Instruction (CPI) & (LIKWID)  A derived metric computed as the quotient of CPU\_CLK\_UNHALTED\_CORE / INSTR\_RETIRED\_ANY to give an estimate of the number of clock cycles per instruction (c.f.~\cite{PattersonHennessy}). \\
\\
Vectorization ratio  & (LIKWID) The ratio of ``packed'' FLOPS to the sum  of all FLOPS. On the KNL platform, these counters may also include integer arithemtic instructions~\cite{LIKWID-website:2020}. \\
    \\
    \texttt{L3CACHE} & (LIKWID) Measures the locality of your data accesses with regard to the L3 cache. It reports the L3 request rate, L3 miss rate, and L3 miss ratio~\cite{LIKWID-website:2020}. (Ivy Bridge only) \\\\
    \end{tabular}
 
\end{table}

%\fix{Need to insert a few sentences here about the counters we are reporting in the study. Yes, there is the table, and it needs to be updated. But we can provide a quick overview here. And we need to explain vector vs. scalar instructions, etc.}

\subsubsection{Testing Procedure}

For each of the different kernels, we execute both \vtk{} and traditional on both the \ivybridge{} and KNL platforms using different datasets (\S\ref{sec:study-data}). 
We are collecting and analyzing the performance counters shown in 
%Appendix \S\ref{sec:appendix-perfCounters}.
Table~\ref{table:perfCounters}.
Of that set, we report a subset in the individual subsections below: total number of instructions executed, number of scalar and vector floating point instructions executed, cycles-per-instruction, vectorization ratio, and L3 cache miss ratio (on the \ivybridge{}) platform.

\subsection{Isocontouring Study}
\label{sec:results-isocontour}

\begin{figure}[t!]
  \centering
  \includegraphics[width=\linewidth]{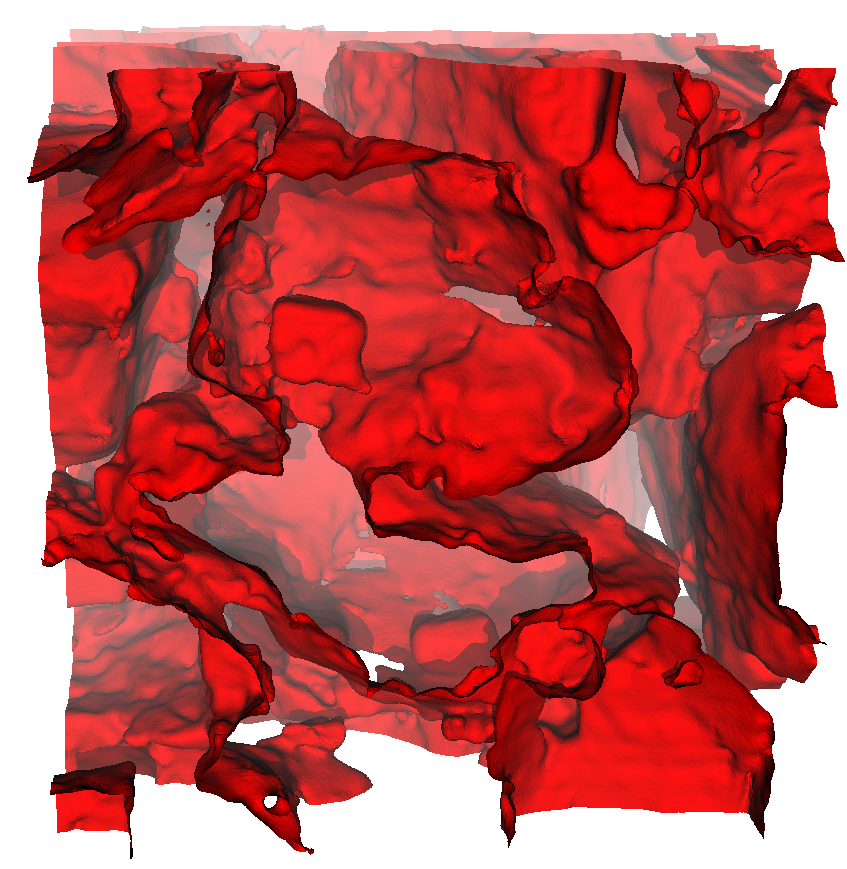}
  \caption{Isocontouring of a $400^3$ subset of smoothed, segmented Sandstone data results in 3,038,366 tris rendered using VisIt.} % v3.0.1.}
  \label{fig:sand400-iso15}
\end{figure}

%Results of isocontour study go here.

\begin{comment}

The test matrix for this study consists of
VTK and \vtkm{} implementations run 
 at varying concurrency on two different platforms:
\ivybridge{}: $P \in (1, 2, 4, 8)$;
KNL: $P \in (1, 2, 4, 8, 16, 32, 64)$.
%
For each test run, we use LIKWID to obtain the hardware performance counters and report a subset of them in Table~\ref{table:isosurface-perfCtrs}. 
\end{comment}

Visual results from this study are shown in Fig.~\ref{fig:sand400-iso15}, and the performance counter and other data appear in Table~\ref{table:isosurface-perfCtrs}.
%
%To begin, we examine the results of comparing the performance of serial versions of both codes. 
%These results are shown in Table~\ref{table:isosurface-perfCtrs}, where we consider runtime and various other hardware performance counters.
%
One observation we draw from the performance data is that for this problem configuration, on the \ivybridge{} platform, the  serial VTK version runs in about 72\% of the time of the serial \vtkm{} version. On the KNL, this difference is more pronounced, with the VTK version running in about 35\% of the time of the \vtkm{} implementation.
This significant difference in runtime is something of a surprise given that both versions are executing the same type of Marching Cubes algorithm, and neither makes use of any special data structures to accelerate the search for cells that containing the isolevel: each implementation must examine each and ever cell in the entire domain, test its nodes for above/below the isolevel, and then generate triangles accordingly. 
%rewquired \vtkm{} version runs about 38\% slower on the \ivybridge{} and about 280\% slower on the KNL platform.

\begin{table}[t!]
\footnotesize{}
\caption{Isocontour Platform and Hardware Performance Counters for the Sandstone $400^3$ Dataset, \ivybridge{} and KNL Platforms. Values close to, but not equal to zero are shown as $\approx{0.00}$. 
Since there is no L3 cache on the KNL, there are no L3 Miss data to report for that platform.}
\label{table:isosurface-perfCtrs}
\begin{center}
\begin{tabular}{@{}llcc@{}}
%\toprule
Counter/Measure                              & Code version & \multicolumn{2}{c}{Platform}                                                                                      \\ 
\midrule

\multirow{4}{*}{Runtime (secs)}      &  \  & \ivybridge{}      & KNL     \\
%\cmidrule(l){3-6}
     & VTK &\multicolumn{1}{r}{$3.97$}& \multicolumn{1}{r}{$5.92$}   \\
     & \vtkm{}   & \multicolumn{1}{r}{$5.48$}  &     \multicolumn{1}{r}{$16.74$}   \\
\multirow{4}{*}{INSTR\_RETIRED\_ANY $10^9$}                   &              &           &                      \\
%\cmidrule(l){3-6}
              & VTK & \multicolumn{1}{r}{$21.96$} &     \multicolumn{1}{r}{$11.10$}          \\        
              & \vtkm{}    & \multicolumn{1}{r}{$22.01$} &     \multicolumn{1}{r}{$21.10$}               \\
 %   \cmidrule(l){3-6}
% ----- 
%\multirow{4}{*}{FLOPS scalar ($*10^9$)}                   &              & \ivybridge{}           & KNL                     \\
\multirow{4}{*}{FLOPS scalar $10^9$}                   &              &           &                     \\
%\cmidrule(l){3-6}
                           & VTK & \multicolumn{1}{r}{$0.08$} &     \multicolumn{1}{r}{$0.50$}          \\
                         & \vtkm{}    & \multicolumn{1}{r}{$0.36$} &     \multicolumn{1}{r}{$0.54$}               \\
 %   \cmidrule(l){3-6}
% ----- 
%    \multirow{4}{*}{FLOPS non-scalar (vector) ($*10^9$)}                   &              &         &              \\
        \multirow{4}{*}{FLOPS non-scalar $10^9$}                   &              &         &              \\
%\cmidrule(l){3-6}
                       & VTK & \multicolumn{1}{r}{$\approx{0.00}$} &     \multicolumn{1}{r}{$0.51$}  \\
                     & \vtkm{}   & \multicolumn{1}{r}{$0.01$} &     \multicolumn{1}{r}{$0.91$}  \\
%    \cmidrule(l){3-6}
% --------------------
% -------                                            
   \multirow{4}{*}{Vectorization \%}                     &              &  &                 \\
%\cmidrule(l){3-3}
                   & VTK & \multicolumn{1}{r}{$\approx{0.00}$}   &  
                                     \multicolumn{1}{r}{$50.50$} \\
                     & \vtkm{}   & \multicolumn{1}{r}{$3.35$} &     \multicolumn{1}{r}{$62.77$}        \\
                                             % \cmidrule(l){1-11} 
        \multirow{4}{*}{CPI}                     &              &            &             \\
%\cmidrule(l){3-6}
                         & VTK & \multicolumn{1}{r}{$0.44$} &   \multicolumn{1}{r}{$0.76$} \\
                         & \vtkm{}   & \multicolumn{1}{r}{$0.55$} & \multicolumn{1}{r}{$1.11$} \\     
\multirow{4}{*}{L3 Miss Ratio \%}                     &              &            &             \\
%\cmidrule(l){3-6}
                         & VTK & \multicolumn{1}{r}{$23.37$} &   N/A \\
                         & \vtkm{}   & \multicolumn{1}{r}{$79.03$} & N/A %\\
%%                                             \cmidrule(l){3-6} 
 % -------                                            
         % \bottomrule
\end{tabular}
\end{center}
\end{table}

To better understand why there is such a performance discrepancy, we turn to the hardware performance counter data shown in Table~\ref{table:isosurface-perfCtrs}.
On the \ivybridge{}, the number of overall number of instructions executed (INSTR\_RETIRED\_ANY) is about the same.
However, the mix of instructions is different between the two implementations, with the \vtk{} code executing about $22\%$ the number of FLOPS as the \vtkm{} version.

For additional clues, we look to the L3 Miss Ratio. The \vtkm{} L3 Miss Ratio is more than $3\times$ that of the \vtk{} code.
This combination, of increased L3 cache misses, which will introduce stalls into the \vtkm{} code, which may or may not be the source of higher CPI. 
We know that the \vtkm{} code consists of sequences of data parallel primitives, where the output of one stage goes on to be the input of the next stage.
We can see the effects of this additional memory handling overhead quite clearly in terms of lower cache utilization, which contribute to the slower runtime and higher CPI.
The runtimes shown here reflect the product of CPI and number of instructions executed~\cite{PattersonHennessy}. 
And the source of higher CPI for the \vtkm{} code is most likely the execution of memory access instructions that are more time consuming.

%Looking at other counters, we see the the \vtkm{} CPI is about 25\% higher than the VTK version, which indicates the mix of instructions being executed by the \vtkm{} implementation is, on the whole, somewhat more time consuming. 
%This situation can happen if one code executes more memory accesses than the other; those executions require more clock cycles to execute than, say, arithmetic operations on register-resident values. 

%but the \vtkm{} CPI is about 25\% higher than the VTK version, most likely due to managing memory as on the KNL. We see what may be additional evidence of this effect in the L3 Miss Ratio, which is about 3.38 times higher in \vtkm{} version. 

On the KNL, we see the \vtkm{} version is executing about $2\times$ the number of instructions, and has a CPI value that is almost $1.5\times$ larger than the VTK version. 
In terms of FLOPS being executed, the number of scalar FLOPS is about the same, but the \vtkm{} code executes almost $2\times$ the number of vector FLOPS. 
Due to limitations of the KNL, we're unable to discern between single-, double-precision and integer vector operations~\cite{LIKWID-website:2020}.
We see the effects of the \vtkm{} increased memory operations directly through the elevated CPI and runtime.

%Since there is no L3 cache on the KNL, and since we know something about the \vtkm{} implementation, we infer that the memory
%This suggests that the KNL version is doing a lot more memory management activities, most likely tasks such as preparing buffers for use by the \vtkm{} worklets. These types of memory intensive operations consume more cycles, which results in the higher CPI. The twice as many instructions consuming about 46\% more time accounts for the difference in runtime between the two versions.

One observation we draw is that use of \vtkm{} and recasting the algorithm to use DPPs comes at a cost, namely less effective cache utilization and a serial runtime that exceeds that of the reference \vtk{} implementation serial runtime. 
This observation is not universally true: for example, Perciano et al., 2020~\cite{Perciano:ICIP16} show that recasting a graphical model optimization problem into \vtkm{} using DPPs results in an implementation that executes far fewer instructions than its C++ counterpart. 
Further studies would include adding finer-grained instrumentation with the LIWKID Marker API into \vtkm{} internals to isolate and study these buffer effects, including factors such as the impact of blocking and chunking the \vtkm{} performs in its back-ends for each device. 

%Where do these extra instructions come from? 
%We speculate 
%One potential source of these extra instructions is the  extra buffer management associated combined with
% the \vtkm{} implementation's use of multiple worklets, although more investigation is needed. 
% combined with the extra buffer management may be the source of the extra work, but more investigation is needed.
%One way to collect evidence needed to answer that question is to do finer-grained instrumentation with the LIKWID Marker API. This activity will make for interesting future work to nail down exactly the source of this ``extra work``. This extra work consists of more instructions being executed, particularly on the KNL, as well as use of instructions that result in more memory intensive operations, which run more slowly and is reflected in CPI on both platforms, and in the L3 Miss Ratio on the KNL.

Another observation is that the Intel compiler is able to achieve high levels of code vectorization on the KNL for both VTK and \vtkm{} implementations, but does not do as well on the \ivybridge{} platform. More investigation is needed to determine why the compiler is having trouble on the \ivybridge{} with these codes. Regardless, the vectorization level does not seem to offer any significant insights into the relative performance differences. 

%For the moment, we can speculate that the \vtkm{} implementation'

%\fix{Need to collect and insert scaling study results here.}

%Show scaling results: runtime, speedup.

%Show some relevant subset of performance counter data to gain understanding about performance.

\subsection{Particle Advection Study}
\label{sec:results-padvect}

\begin{figure}[t!]
\centering
\includegraphics[width=0.9\linewidth]{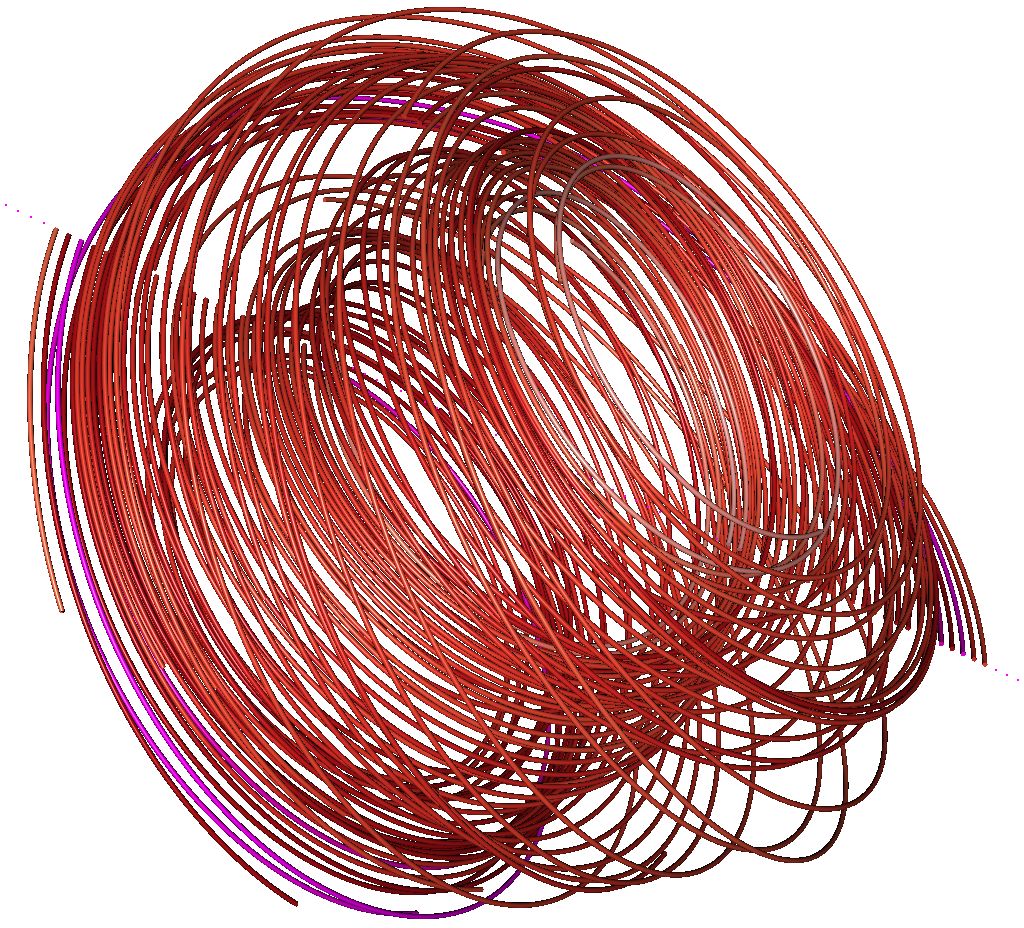}
\caption{Streamline results from the particle tracing method run using the NIMROD dataset rendered using VisIt.
%Since particle advection performance is dependent in part upon the underlying vector field and placement of seed points, we use three different datasets in our performance study: astrophysics simulation (left), thermal hydraulics (middle), and magnetic fusion modeling (right). This diversity of datasets help to even out data dependent performance characteristics in the experimental results.
%\fix{Update caption since we are using only nimrod data.}
}
\label{fig:padvect-samples}
\end{figure}

%Particle advection study results go here.

\begin{comment}

The test matrix for this study consists of
\emph{which implementations} run 
 at varying concurrency on two different platforms:
\ivybridge{}: $P \in (1, 2, 4, 8)$;
KNL: $P \in (1, 2, 4, 8, 16, 32, 64)$.
For each test run, we use LIKWID to obtain the hardware performance counters and report a subset of them in \emph{Table XX, which doesn't exist yet.}

\end{comment}

Visual results from this study are shown in Fig.~\ref{fig:padvect-samples}, and the performance counter and other data appear in Table~\ref{table:padvect-perfCtrs}. 
%Runtimes and performance counters from serial versions of the VTK and \vtkm{} implementations appear in Table~\ref{table:padvect-perfCtrs}. 
From this data, we see that the \vtkm{} implementation runs faster on both platforms; on the \ivybridge{}, it runs in about 89\% of the VTK method, and on the KNL, the difference is more pronounced, where it runs in about 26\% of the VTK method. 

On the \ivybridge{}, we see the \vtkm{} implementation executes far fewer total instructions (INSTR\_RETIRED\_ANY), only about 23\% as many as the VTK method. 
Proportionally speaking, the \vtk{} code is executing a much greater percentage of non-floating point instructions as evidenced by comparing the ratio of FLOPS/INSTR\_RETIRED\_ANY between the two. 
These extra non-floating point instructions are most likely due to extensive bounds and indexing calculations: as a particle is advected, the next step of the algorithm is to determine in which mesh cell the particle is contained. 
%The \vtkm{} implementation is also executing far fewer scalar and vector floating point instructions, less than half as many, as the VTK implementation.

We know something about the \vtkm{} implementation: it is a recasting of the particle advection into a series of data parallel primitives, where output from one stage is input to the next stage.
And as we saw with the isocontouring study, this type of design pattern can be result in less efficient use of the memory hierarchy. We see evidence of this effect when looking at the L3 Miss Ratio in Table~\ref{table:padvect-perfCtrs}, as well as at the significantly higher CPI, which reflects the impact of costly non-cache memory accesses. 

%On the other hand, the \vtkm{} CPI is significantly larger, nearly 4 times as large. 
%It appears that on the \ivybridge{} platform, the \vtkm{} implementation is executing proportionally many more instructions that are costly, such as managing and filling memory buffers, which also results in higher levels of L3 cache misses.
%Despite the significant differences in amount of overall and numerical instructions being executed by the two, about $5\times$ and $2\times$ respectively, the difference in CPI going the other way causes the total runtime difference to be much smaller than one might expect looking at only instruction count. 

On the KNL platform, the runtime difference between the two is even more pronounced. 
Again, we see the \vtk{} implementation executes more overall (about $2.5\times$) and numerical instructions (between  about $3.1\times$ and $3.3\times$).
We believe the increased CPI on the \knl{} for the \vtk{} implementation to be a result of less efficient cache utilization: particle advection memory accesses are unstructured, so there is little opportunity for speculative memory fetches.
The \vtkm{} code reorders memory accesses to fit its DPP execution pattern, and in this case, appears to have better cache utilization as shown by a lower CPI.
%
%In contrast to the \ivybridge{} platform, we see the CPI level reversed on the KNL, with the most likely explanation being that the VTK method was able to take better advtange of the large LLC on the \ivybridge{}. Since the LLC on the KNL is much smaller (30 MiB vs. 1 MiB), the VTK CPI is adversely affected. 
For the \vtkm{} implementation, the combination of fewer instructions and lower CPI combine to produce a dramatically lower runtime.

%As we have seen in other studies (c.f.,~\cite{Perciano:2020:ISC}), the process of refactoring an algorithm to use DPPs can produce a significant reduction in the total number of instructions executed. 

% Table~\ref{table:perfCounters}.

\begin{table}[t!]
\footnotesize{}
\caption{Particle Advection Platform and Hardware Performance Counters for the Nimrod, \ivybridge{} and KNL Platforms. Values close to, but not equal to zero are shown as $\approx{0.00}$, while values equal to zero are shown as $=0.00$. 
Since there is no L3 cache on the KNL, there are no L3 Miss data to report for that platform.
%On the \ivybridge{}, the last level cache (LLC) is L3, and on the KNL, is L2. \fix{data now copied need, next need to do analysis and writeup.} 
}
\label{table:padvect-perfCtrs}
\begin{center}
\begin{tabular}{@{}llcc@{}}
%\toprule
Counter/Measure                              & Code version & \multicolumn{2}{c}{Platform}                                                                                      \\ 
\midrule

\multirow{4}{*}{Runtime (secs)}              &               & \ivybridge{}            & KNL     \\
%\cmidrule(l){3-6}
     & VTK &\multicolumn{1}{r}{$7.68$}& \multicolumn{1}{r}{$7.72$}   \\
     & \vtkm{}   & \multicolumn{1}{r}{$6.81$}  &     \multicolumn{1}{r}{$2.03$}   \\
\multirow{4}{*}{INSTR\_RETIRED\_ANY $10^9$}                   &              &           &                      \\
%\cmidrule(l){3-6}
              & VTK & \multicolumn{1}{r}{$25.56$} &     \multicolumn{1}{r}{$5.92$}          \\        
              & \vtkm{}    & \multicolumn{1}{r}{$5.91$} &     \multicolumn{1}{r}{$2.34$}               \\
 %   \cmidrule(l){3-6}
% ----- 
%\multirow{4}{*}{FLOPS scalar ($*10^9$)}                   &              & \ivybridge{}           & KNL                     \\
\multirow{4}{*}{FLOPS scalar $10^9$}                   &              &           &                     \\
%\cmidrule(l){3-6}
                           & VTK & \multicolumn{1}{r}{$0.27$} &     \multicolumn{1}{r}{$0.85$}          \\
                         & \vtkm{}    & \multicolumn{1}{r}{$0.11$} &     \multicolumn{1}{r}{$0.26$}               \\
 %   \cmidrule(l){3-6}
% ----- 
%    \multirow{4}{*}{FLOPS non-scalar (vector) ($*10^9$)}                   &              &         &              \\
        \multirow{4}{*}{FLOPS non-scalar $10^9$}                   &              &         &              \\
%\cmidrule(l){3-6}
                       & VTK & \multicolumn{1}{r}{$0.01$} &     \multicolumn{1}{r}{$0.68$}  \\
                     & \vtkm{}   & \multicolumn{1}{r}{$=0.00$} &     \multicolumn{1}{r}{$0.22$}  \\
%    \cmidrule(l){3-6}
% --------------------
% -------                                            
   \multirow{4}{*}{Vectorization \%}                     &              &  &                 \\
%\cmidrule(l){3-3}
                   & VTK & \multicolumn{1}{r}{$2.73$}   &  
                                     \multicolumn{1}{r}{$44.52$} \\
                     & \vtkm{}   & \multicolumn{1}{r}{$=0.00$} &     \multicolumn{1}{r}{$45.75$}        \\
                                             % \cmidrule(l){1-11} 
        \multirow{4}{*}{CPI}                     &              &            &             \\
%\cmidrule(l){3-6}
                         & VTK & \multicolumn{1}{r}{$0.74$} &   \multicolumn{1}{r}{$1.94$} \\
                         & \vtkm{}   & \multicolumn{1}{r}{$2.78$} & \multicolumn{1}{r}{$1.28$} \\     
\multirow{4}{*}{L3 Miss Ratio \%}                     &              &            &             \\
%\cmidrule(l){3-6}
                         & VTK & \multicolumn{1}{r}{$33.53$} &   \multicolumn{1}{r}{N/A} \\
                         & \vtkm{}   & \multicolumn{1}{r}{$57.09$} & \multicolumn{1}{r}{N/A} %\\  
%%                                             \cmidrule(l){3-6} 
 % -------                                            
%        \\ \bottomrule
\end{tabular}
\end{center}
\end{table}

\begin{comment}

\begin{figure*}[t!]
\centering
\includegraphics[width=0.32\textwidth]{images/astro0000.jpeg}
\includegraphics[width=0.32\textwidth]{images/fishtank0000.jpeg}
\includegraphics[width=0.32\textwidth]{images/nimrod0000.jpeg}
\caption{Since particle advection performance is dependent in part upon the underlying vector field and placement of seed points, we use three different datasets in our performance study: astrophysics simulation (left), thermal hydraulics (middle), and magnetic fusion modeling (right). This diversity of datasets help to even out data dependent performance characteristics in the experimental results.
}
\label{fig:padvect-samples}
\end{figure*}

\end{comment}

\subsection{Stencil-based Smoothing Study}
\label{sec:results-stencil}

The test matrix in this portion of the study consists of three different implementations of the stencil-based smoothing (traditional \cppomp{}, \vtkmFM{}/OpenMP, and \vtkmPN{}/OpenMP) run at varying concurrency on two different platforms:
\ivybridge{}: $P \in (1, 2, 4, 8)$;
KNL: $P \in (1, 2, 4, 8, 16, 32, 64)$.
For each test run, we use LIKWID to obtain the hardware performance counters and report them in Table~\ref{table:stencil-perfCtrs}.

\begin{figure}[t!]
\includegraphics[width=0.98\linewidth]{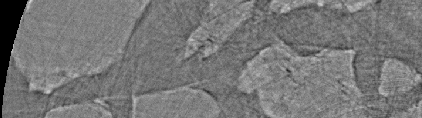}
\includegraphics[width=0.98\linewidth]{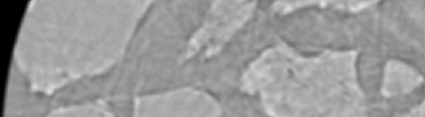}
\includegraphics[width=0.98\linewidth]{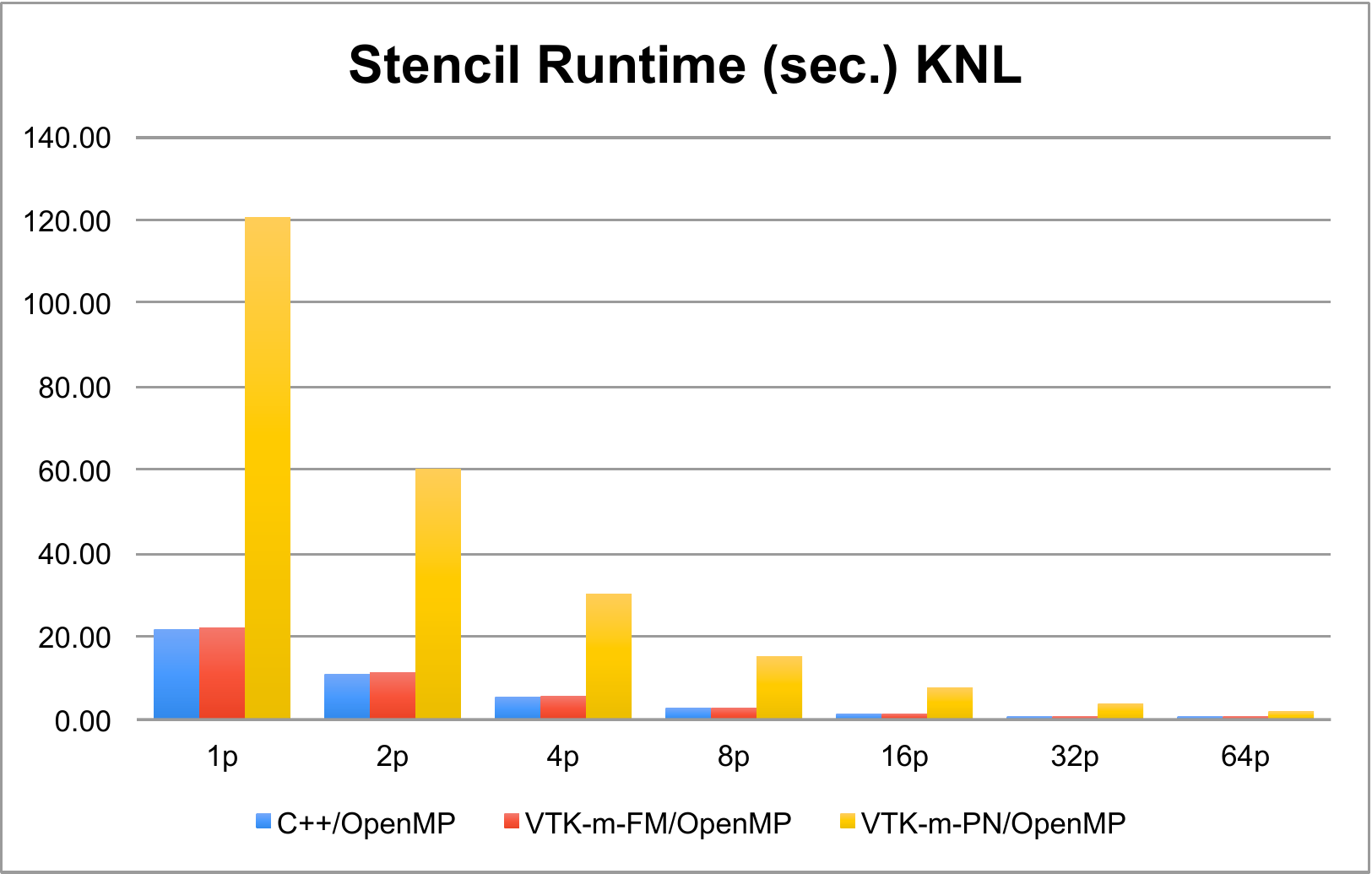}
\includegraphics[width=0.98\linewidth]{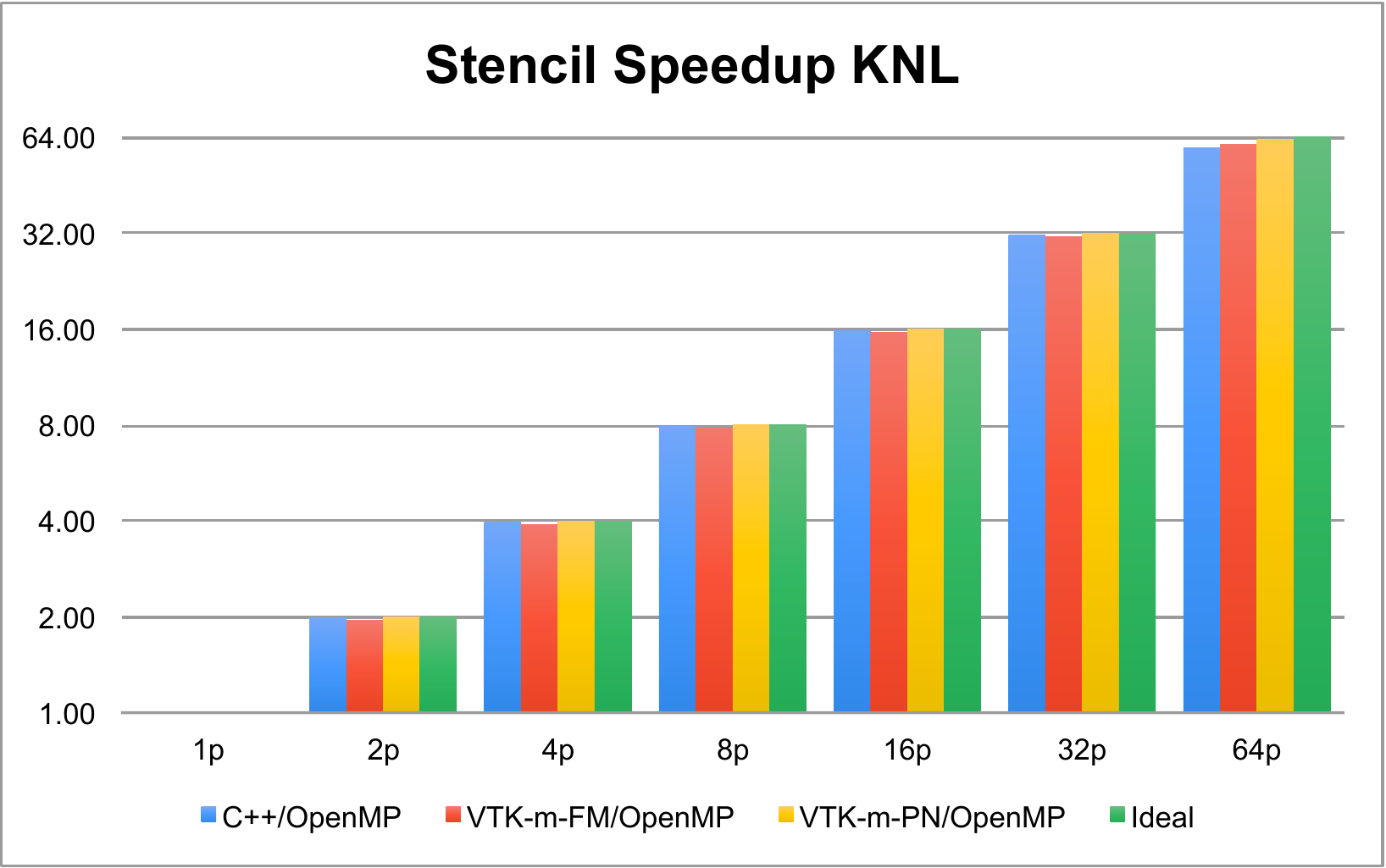}
\caption{(Top) Original image data of a geological sample, obtained by micro computed tomography at the Advanced Light Source. (Middle-top) Results of Guassian smoothing using  the \cppomp{} implementation with $5x5$ stencil. (Bottom 2 images) Absolute runtime and speedup of three stencil implementations on the KNL platform at varying concurrency with a $19x19$ stencil on a high-resolution ($5K \times 5K$) input image.}
\label{fig:sandstone-stencil}
\end{figure}

Since we are doing a strong-scaling configuration in these tests, we expect the runtime to drop with increasing concurrency.
We see such behavior in Fig.~\ref{fig:sandstone-stencil}, where absolute runtime increases at varying concurrency, as well as speedup compared to serial. 
In this case, all algorithms exhibit near-perfect scaling characteristics. This result is expected since this algorithm is ``embarrassingly parallel'': the computation of each output pixel is completely independent of the computation at all other output pixels, and these computations may be performed in parallel. 

While it is the case all methods exhibit near-perfect scaling, we see that the \vtkmPN{}/OpenMP method has runtime far greater than the other two methods. 
For evidence of why this is the case, we turn to hardware performance counter data, which we present in Table~\ref{table:stencil-perfCtrs}.
The performance data in this table is from a serial run of each method.
What is not shown in the table is data from varying concurrency runs. It turns out that these values do not change in any significant way with increasing concurrency on this problem with these codes.

The first, and most significant, observation is that \vtkmPN{} executes about 5 times as many instructions as the other implementation, which would account for its 5-fold increase in runtime compared to the other methods.  
This result is most likely due to the effort on the part of VTK-m to prepare point neighborhood collections of data from the original input data.
In contrast, the other two methods do their own indexing into the original data array, an approach that appears to result in far fewer instructions being executed.

The second observation is that the \vtkmPN{} method executes far fewer ``packed'', or vector, floating point instructions than the other two methods.
This result is most likely due to the fact that the \vtkmPN{} worklet is accessing data through a method, e.g., \texttt{getData(i, j, k)}, rather than via an index that is a function of a loop variable, e.g., \texttt{data[i+offset]}. 

A third observation is that the \cppomp{} implementation appears to have a very slight performance edge compared to the \vtkmFM{} approach in terms of runtime, number of instructions executed, and L3 cache utilization.
This result is due to the differences in the way these two methods are parallelized. 
In the case of \cppomp{}, each OpenMP ``thread'' processes an entire row of pixels; OpenMP parallelism happens at the outer loop level, resulting in a somewhat coarse level of parallelism. 
In the case of the \vtkmFM{} method, VTK-m parallelizes the call to the worklet, which is invoked once per output pixel. Therefore, \vtkmFM{} is operating at a finer-grained level of parallelism, compared to the \cppomp{}.
The resulting performance data suggest that the \cppomp{} approach, with its coarser-grained approach to parallelism, runs a bit faster most likely due to its better use of temporal cache locality, as evidenced by the L3 cache data on the IvyBridge platform. 

\begin{table}[h!]
\footnotesize{}
\caption{Stencil Convolution Platform and Hardware Performance Counters for the Sandstone5K Dataset, $19 \times 19$ stencil size, \ivybridge{} and KNL Platforms.}
\label{table:stencil-perfCtrs}
\begin{center}
\begin{tabular}{@{}llcc@{}}
%\toprule
Counter/Measure                              & Code version & \multicolumn{2}{c}{Platform}                                                                                      \\ 
\midrule

\multirow{4}{*}{Runtime (secs)}              &               & \ivybridge{}            & KNL     \\
%\cmidrule(l){3-6}
     & C++ &\multicolumn{1}{r}{$7.59$}& \multicolumn{1}{r}{$21.65$}   \\
     & \vtkmFM{}   & \multicolumn{1}{r}{$8.33$}  &     \multicolumn{1}{r}{$22.09$}                \\
     & \vtkmPN{}     & \multicolumn{1}{r}{$37.35$}  &     \multicolumn{1}{r}{$120.49$}                     \\
%     \cmidrule(l){3-6}
% -------------------
%\multirow{4}{*}{INSTR\_RETIRED\_ANY ($*10^9$)}                   &              & \ivybridge{}           & KNL                     \\
%\multirow{4}{*}{INSTR\_RETIRED\_ANY ($*10^9$)}                   &              &           &                      \\
%\multirow{4}{*}{INSTR\_RETIRED\_ANY}                   &              &           &                      \\
\multirow{4}{*}{INSTR\_RETIRED\_ANY $10^9$}                   &              &           &                      \\
%\cmidrule(l){3-6}
                              & C++ & \multicolumn{1}{r}{$57.60$} &     \multicolumn{1}{r}{$41.51$}          \\
                             & \vtkmFM{}    & \multicolumn{1}{r}{$59.60$} &     \multicolumn{1}{r}{$42.59$}               \\
                             & \vtkmPN{} & \multicolumn{1}{r}{$249.37$} &
                           \multicolumn{1}{r}{$215.59$}  \\
 %   \cmidrule(l){3-6}
% ----- 
%\multirow{4}{*}{FLOPS scalar ($*10^9$)}                   &              & \ivybridge{}           & KNL                     \\
\multirow{4}{*}{FLOPS scalar $10^9$}                   &              &           &                     \\
%\cmidrule(l){3-6}
                           & C++ & \multicolumn{1}{r}{$5.76$} &     \multicolumn{1}{r}{$4.47$}          \\
                         & \vtkmFM{}    & \multicolumn{1}{r}{$5.77$} &     \multicolumn{1}{r}{$4.47$}               \\
                     & \vtkmPN{} & \multicolumn{1}{r}{$30.24$} &
                           \multicolumn{1}{r}{$20.81$}  \\
 %   \cmidrule(l){3-6}
% ----- 
%    \multirow{4}{*}{FLOPS non-scalar (vector) ($*10^9$)}                   &              &         &              \\
        \multirow{4}{*}{FLOPS non-scalar $10^9$}                   &              &         &              \\
%\cmidrule(l){3-6}
                       & C++ & \multicolumn{1}{r}{$8.68$} &     \multicolumn{1}{r}{$14.93$}  \\
                     & \vtkmFM{}   & \multicolumn{1}{r}{$8.67$} &     \multicolumn{1}{r}{$14.99$}  \\
                     & \vtkmPN{} & \multicolumn{1}{r}{$=0.00$} &
                           \multicolumn{1}{r}{$1.16$}  \\
%    \cmidrule(l){3-6}
% --------------------
% -------                                            
   \multirow{4}{*}{Vectorization \%}                     &              &  &                 \\
%\cmidrule(l){3-3}
                   & C++ & \multicolumn{1}{r}{$60.11$}   &  
                                     \multicolumn{1}{r}{$76.97$} \\
                     & \vtkmFM{}   & \multicolumn{1}{r}{$60.05$} &     \multicolumn{1}{r}{$77.04$}        \\
                 & \vtkmPN{}  & \multicolumn{1}{r}{$=0.00$}  &  \multicolumn{1}{r}{$5.27$}  \\ 
                                             % \cmidrule(l){1-11} 
     \multirow{4}{*}{CPI}                     &              &            &             \\
%\cmidrule(l){3-6}
                         &  C++ & \multicolumn{1}{r}{$0.32$} &   \multicolumn{1}{r}{$0.76$} \\
                         & \vtkmFM{}    & \multicolumn{1}{r}{$0.34$} & \multicolumn{1}{r}{$0.77$} \\  
                    & \vtkmPN{}    & \multicolumn{1}{r}{$0.37$} & \multicolumn{1}{r}{$0.83$} \\           
\multirow{4}{*}{L3 Miss Ratio \%}                     &              &            &             \\
%\cmidrule(l){3-6}
                         & C++ & \multicolumn{1}{r}{$15.85$} &   N/A \\
                         & \vtkmFM{}   & \multicolumn{1}{r}{$37.14$} & N/A \\  
                         & \vtkmPN{}     & \multicolumn{1}{r}{$47.84$} &    N/A % \\ 
%%                                             \cmidrule(l){3-6} 
 % -------                                            
  %      \\ \bottomrule
\end{tabular}
\end{center}
\end{table}

\subsection{Discussion of Results}

One observation from these studies is that the hardware performance counters are useful for understanding more about why two different methods have different runtimes.
In some cases, they are executing different absolute numbers of instructions. In other cases, the type of instruction being executed can take more time: instructions that load/store memory can take significantly longer than simple arithmetic instructions.
In other cases, demographics of compiler-generated scalar or vector arithmetic instructions can impact overall runtime.

The three cases we present all exhibit different aspects of why a method might have better or worse runtime than another.
In some cases, the way an algorithm is implemented, such as VTK vs. \vtkm{}, can have a dramatic impact on overall number of instructions, a fact that is corroborated by other recent studies (c.f.,~\cite{Perciano:2020:ISC}).
In other cases, the  buffer management needed to implement a complex, multi-stage processing pipeline may trigger more memory movement instructions, which may be more expensive and result in higher CPI values, and we see evidence of this in two of the examples. 

\begin{comment}

Discuss results.

What insights did we gain? Did we answer the research questions?

Mention why we focused only on CPU stuff even though VTK-m will run on the GPU.

As we have seen in other studies , the process of refactoring an algorithm to use DPPs can produce a significant reduction in the total number of instructions executed.

Why only serial for isocontour and padvect?

\end{comment}

\section{Conclusions and Future Work}

This study demonstrates the value of using a methodology to collect and analyze hardware performance counter data to better understand the reasons behind performance differences in multiple implementations of three kernels common in visualization and analysis processing: isocontouring, particle advection, and stencil-based convolution.
While runtime alone is useful, having  performance counter data enables analysis at a deeper level. 
Although this approach of using hardware counters to do performance analysis
%a deeper introspection on code performance 
is not new, it is relatively new to the visualization community, where previous performance studies have focused primarily on using runtime and runtime derivatives as the performance measure. 

While the stencil-based computation work includes a modest strong scaling study, our results focus primarily on comparing serial implementations of methods implemented in VTK and \vtkm{}. 
While \vtkm{} is intrinsically (shared memory) parallel, the VTK-based implementations are inherently serial due to limitations inside VTK itself.
Future work may include finding workarounds to VTK limitation so that it is possible to do more widespread performance comparison studies across a larger set of key \vtk{} and \vtkm{} algorithms.

While our work focuses on a multi-core CPU platform, \vtkm{} is capable of emitting code that can run on a GPU. According to the OpenMP standard~\cite{OpenMP-GPU:2020}, OpenMP is capable of emitting GPU device code  (called ``device offload''), however there are significant constraints on exactly what may be successfully offloaded to the GPU.
While it is likely possible to do a direct custom C++/OpenMP comparison with \vtkm{} implementation, doing so with an C++/OpenMP application that is engineered to invoke VTK methods may not be feasible, due to OpenMP device offload limitations.

Additional future work will include doing finer grained studies that isolate and quantify the cost of buffer and memory management needed for \vtkm{} and DPP-based patterns. One dimension of that future work would include studies that examine the performance impact of changing the blocking factor used by the \vtkm{} back-end as it divides up data for worklets to execute in parallel.
Similarly, while this study asks some preliminary questions for each of these three methods and their multiple implementations, a significant amount of in-depth analysis work remains to fully characterize performance, such as analyzing memory system utilization and its relationship to data management strategies inside the code.

%identifying the source of ``extra instructions'' we observe in our studies as we compare performance of different implementations. These extra instructions are most likely associated with managing memory buffers internal to \vtkm{} as part of complex, multi-stage processing pipelines. One approach for pursuing this study could be finer-grained instrumentation using LIKWID's Marker API. 

The methodology we present in our studies is useful for other applications, particularly those where you might not have the source code. LIKWID can be used in a fashion that does not require any coding instrumentation at all. In that case, the counters are collected over the entire application run, including I/O and other operations that may not or may not be of interest.  

\bibliographystyle{IEEEtran}
\bibliography{IEEEabrv,main.bib}
%\begin{thebibliography}{1}

%\end{thebibliography}

%\clearpage
%\appendix
%\input{appendix-perfCounters}

\end{document}